\documentclass[sigplan,screen,nonacm]{acmart} 

\AtBeginDocument{%
  }

\setcopyright{acmlicensed}
\copyrightyear{2018}
\acmYear{2018}
\acmDOI{XXXXXXX.XXXXXXX}
\acmConference[Conference acronym 'XX]{Make sure to enter the correct
  conference title from your rights confirmation email}{June 03--05,
  2018}{Woodstock, NY}
\acmISBN{978-1-4503-XXXX-X/2018/06}

\begin{document}

\title{A Multi-Scale Attention-Enhanced Architecture for Gravity Wave Localization in Satellite Imagery}

\author{Seraj Al Mahmud Mostafa}
\affiliation{
  \institution{Department of Information Systems\\University of Maryland, Baltimore County}
  \city{Baltimore}
  \state{Maryland}
  \country{USA}
}


\author{Jianwu Wang}
\affiliation{
  \institution{Department of Information Systems\\University of Maryland, Baltimore County}
  \city{Baltimore}
  \state{Maryland}
  \country{USA}
}






\renewcommand{\shortauthors}{Seraj Mostafa et al.}

\begin{abstract}
Satellite images present unique challenges due to their high object variability and lower spatial resolution, particularly for detecting atmospheric gravity waves which exhibit significant variability in scale, shape, and pattern extent, making accurate localization highly challenging. This variability is further compounded by dominant unwanted objects such as clouds and city lights, as well as instrumental noise, all within a single image channel, while conventional detection methods struggle to capture the diverse and often subtle features of gravity waves across varying conditions. To address these issues, we introduce YOLO-DCAT incorporating Multi Dilated Residual Convolution (MDRC) and Simplified Spatial and Channel Attention (SSCA), an enhanced version of YOLOv5 specifically designed to improve gravity wave localization by effectively handling their complex and variable characteristics. MDRC captures multi-scale features through parallel dilated convolutions with varying dilation rates, while SSCA focuses on the most relevant spatial regions and channel features to enhance detection accuracy and suppress interference from background noise. In our experiments, the improved model outperformed state-of-the-art alternatives, improving mean Average Precision (mAP) by over 14\% and Intersection over Union (IoU) by approximately 17\%, demonstrating significantly improved localization accuracy for gravity waves in challenging satellite imagery and contributing to more precise climate research and modeling.

\end{abstract}



\keywords{Localization, Detection, YOLO, Gravity Wave, Attention, Dilated Convolution, Noisy Satellite Data.}


\maketitle

\section{Introduction}
Satellite data plays a vital role in Earth informatics by offering valuable insights into the planet's environment and climate, providing critical information for interpreting a range of phenomena such as atmospheric conditions, sea surface changes, and environmental shifts. Additionally, satellite imagery is indispensable for studying gravity waves in the Earth's atmosphere. Gravity waves are oscillations caused by buoyancy forces that are distinct from gravitational waves and play a crucial role in atmospheric dynamics and energy transfer \citep{jovanovic2018nature}. They significantly impact weather patterns, atmospheric composition, and climate systems, making their study essential for comprehensive climate understanding and modeling \citep{alexander1997model}.

Localizing gravity waves in the atmosphere is vital for climate research, as it helps quantify their effects on global atmospheric circulation and energy distribution \citep{fritts2003gravity}. However, despite the wealth of information provided by satellite datasets, they present significant complexities \cite{guo2007semantic}. Particularly, the satellite data poses substantial difficulties in gravity wave datasets, where various interferences, such as city lights, clouds, and instrumental noise obscure subtle atmospheric phenomena \cite{mostafa2025gwavenet, gonzalez2022atmospheric}. These datasets also encounter unavoidable challenges including significant variability in the scale, shape, and extent of the main object patterns. Gravity waves can be mixed with or hindered by interference from occlusion and overlap, complicating the detection process. Moreover, the fact that gravity wave datasets are captured and stored in a single band makes it challenging to distinguish and remove unwanted objects. The relative infrequency and low significance of gravity waves in these datasets further complicate their detection \citep{gravity_wave_data}.

To tackle these challenges and improve gravity wave localization, we propose YOLO-DCAT (YOLO with \underline{D}ilated \underline{C}onvolution and \underline{A}ttention-aided \underline{T}echnique), which provides enhancements to the existing state-of-the-art YOLOv5 object detection model \cite{glenn_jocher_2020_4154370}. Our approach focuses on two key contributions as follows: \textbf{1) Multi Dilated Residual Convolution (MDRC):} This modification to the network's backbone increases the receptive field without losing spatial resolution. The multi-dilation approach applies different dilation rates to convolutional layers, enabling the capture of features at various scales simultaneously. The residual approach \cite{he2016deep} with skip connections allows the network to learn residual functions, facilitating gradient flow and improving training efficiency. Feature fusion combines information from different dilation rates, enhancing the model's ability to detect variable-scale gravity wave patterns effectively. \textbf{2) Simplified Spatial and Channel Attention (SSCA):} Implemented after each Dilated Residual Block in the backbone, this mechanism enhances important features while suppressing noise interference. It focuses on relevant spatial areas and channels simultaneously, improving the model's ability to distinguish gravity waves from background interference and unwanted objects.

\begin{figure*}[htb!]
    \centering
    \begin{tabular}{c@{\hspace{1mm}}c@{\hspace{1mm}}c@{\hspace{1mm}}c@{\hspace{1mm}}c@{\hspace{1mm}}c}
        \small\textbf{Labeled Data} & \small\textbf{Y.+MDRC+SSCA} & \small\textbf{CBAM \cite{woo2018cbam}} & \small\textbf{ViT \cite{dosovitskiy2020image}} & \small\textbf{Transformer \cite{vaswani2017attention}} & \small\textbf{YOLO \cite{glenn_jocher_2020_4154370}} \\ 
        \includegraphics[width=0.16\linewidth, height=3cm]{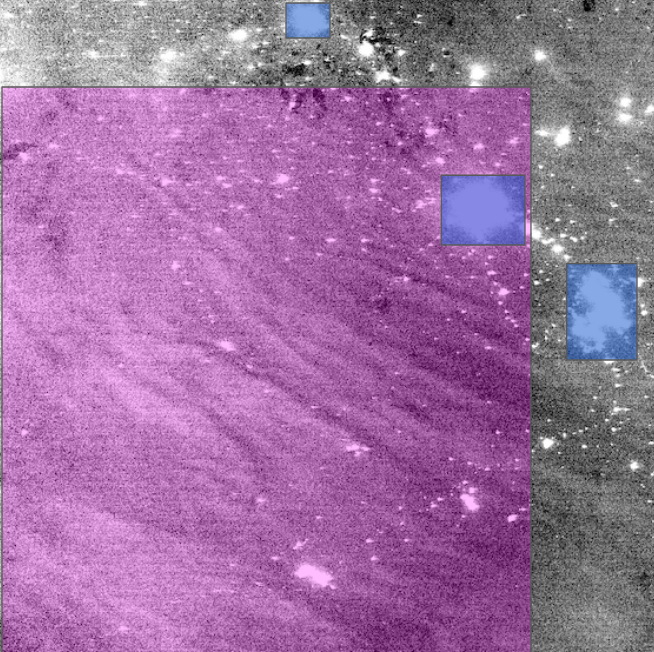} &
        \includegraphics[width=0.16\linewidth, height=3cm]{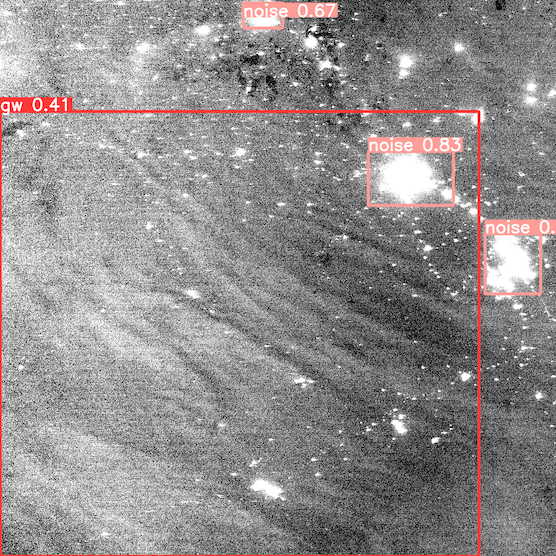} &
        \includegraphics[width=0.16\linewidth, height=3cm]{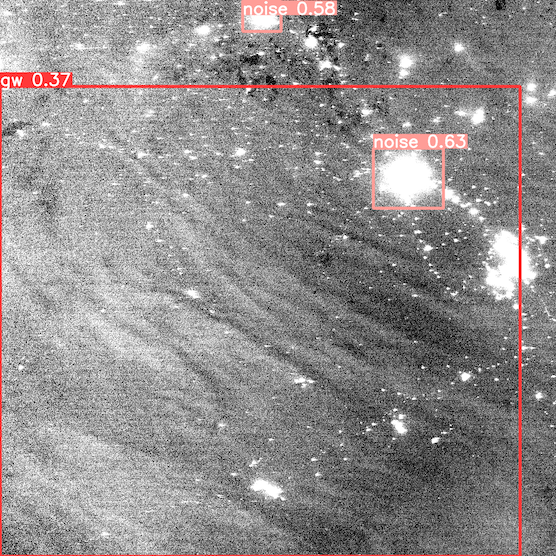} &
        \includegraphics[width=0.16\linewidth, height=3cm]{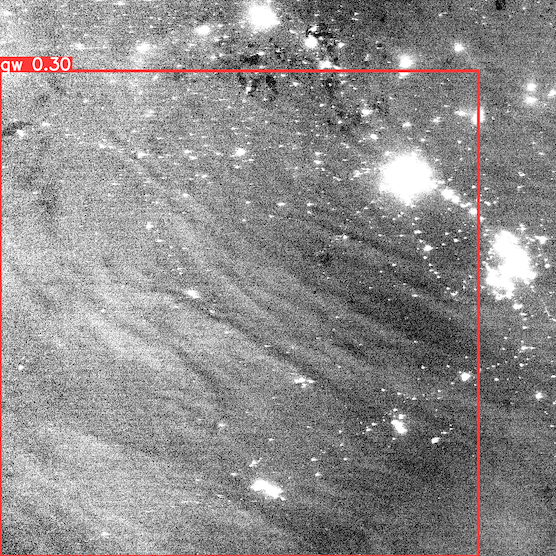} &
        \includegraphics[width=0.16\linewidth, height=3cm]{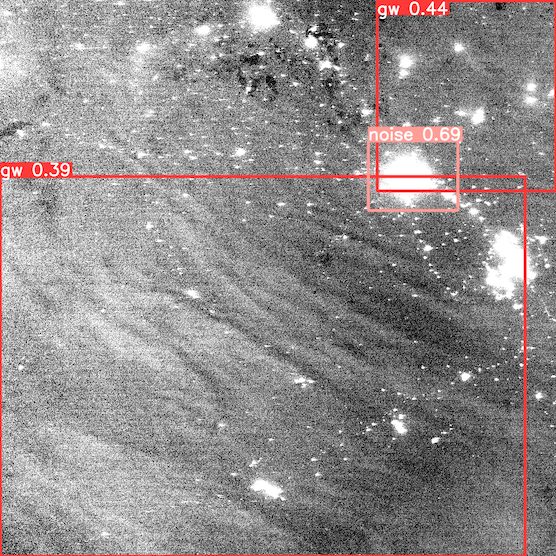} &
        \includegraphics[width=0.16\linewidth, height=3cm]{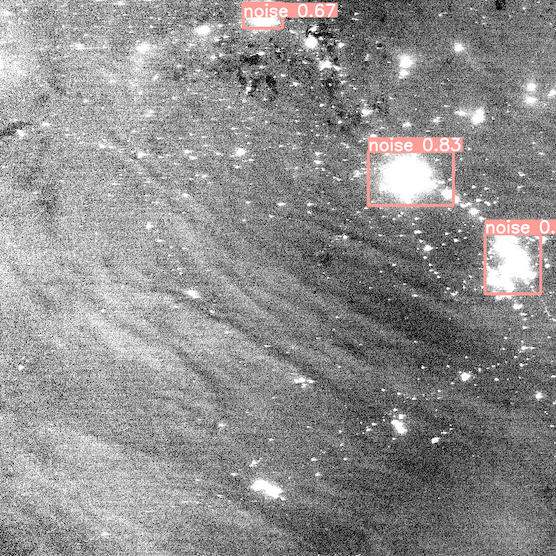} \\ 
        [1ex]
        \includegraphics[width=0.16\linewidth, height=3cm]{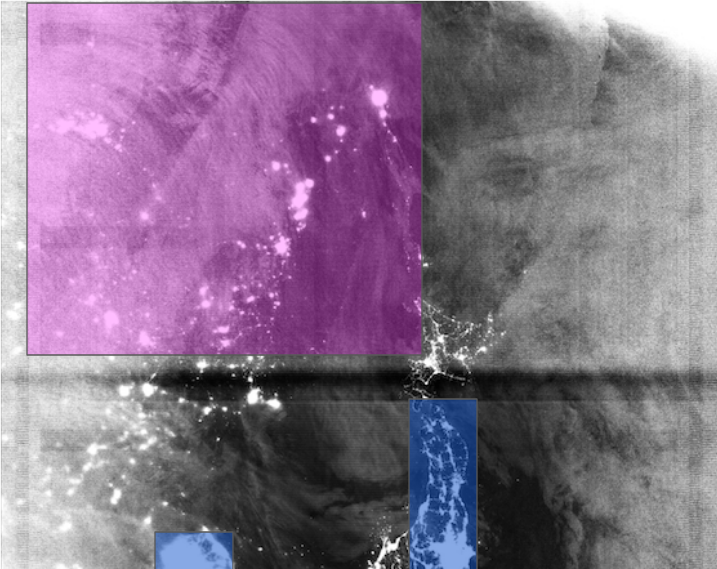} &
        \includegraphics[width=0.16\linewidth, height=3cm]{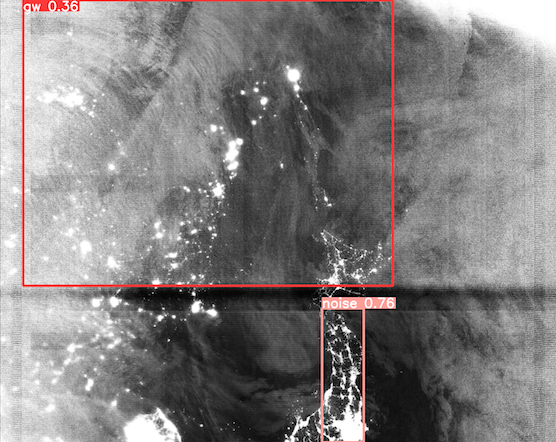} &
        \includegraphics[width=0.16\linewidth, height=3cm]{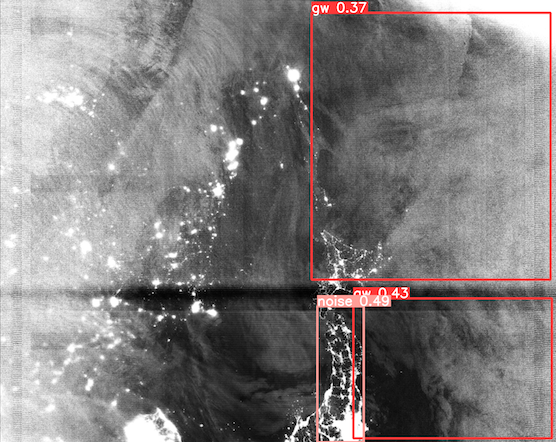} &
        \includegraphics[width=0.16\linewidth, height=3cm]{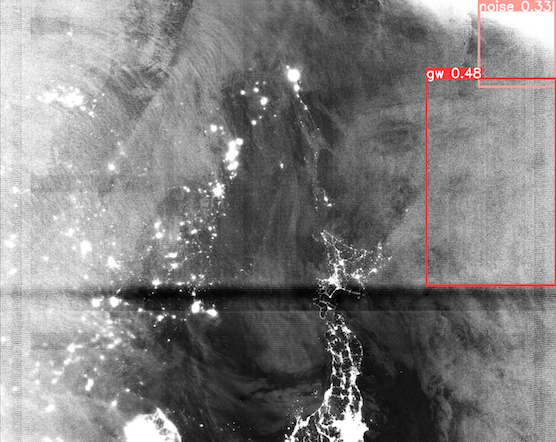} &
        \includegraphics[width=0.16\linewidth, height=3cm]{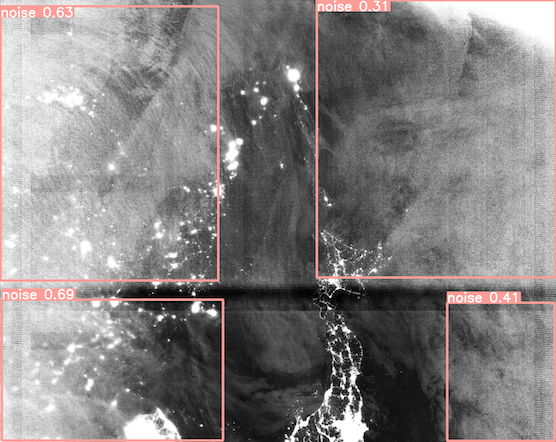} &
        \includegraphics[width=0.16\linewidth, height=3cm]{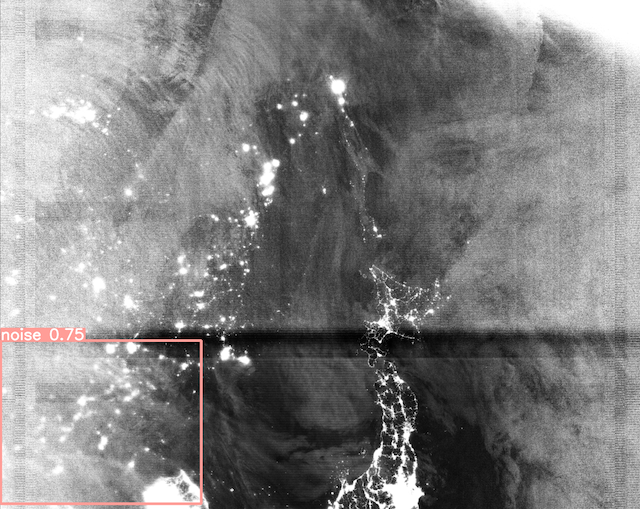}
    \end{tabular}
    \caption{Comparing the localization of Gravity Waves using various baseline and state-of-the-art models, including the proposed YOLO+MDRC+SSCA (denoted as Y.+MDRC+SSCA in the second column) model. The leftmost column represents the labeled data, where the object of interest (Gravity Waves) is highlighted in pink shades, while non-objects of interest (such as city lights, and clouds) are highlighted in blue shades. The top row displays data with more visible gravity wave occurrences, whereas the bottom row presents more challenging scenarios where gravity waves are partially obscured, blurred, or embedded within clouds. (Please refer to supplementary material for more results).}
    \label{fig:vis-res}
\end{figure*}

These enhancements enable our model to better handle the challenges of noisy satellite data and improve the accuracy of gravity wave localization, contributing to more precise climate research and modeling. In Figure \ref{fig:vis-res}, we present results demonstrating the effectiveness of our proposed model compared to other state-of-the-art models. The detection results show that our approach delivers more promising performance than both baseline and state-of-the-art models, with improved localization accuracy and higher confidence scores. A detailed discussion of the experimental results is provided in Section \ref{sec:experiments}.

The structure of this paper is organized as follows. Section~\ref{sec:background} provides an overview of the Gravity Wave dataset and the YOLO model architecture. Section~\ref{sec:relatedwork} reviews related work in dilated convolutions and attention mechanisms. In Section~\ref{sec:methodology}, we present the proposed methodology with detailed explanations of MDRC and SSCA components, followed by the experimental setup and comprehensive results in Section~\ref{sec:experiments}. Section~\ref{sec:disc} offers a thorough discussion of our findings and their implications, and finally, Section~\ref{sec:conclusions} concludes this study with future research directions.

\section{Background}
\label{sec:background}

\subsection{Gravity Wave Dataset}
This study utilizes data from the night band of the Visible Infrared Imaging Radiometer Suite (VIIRS) Day/Night Band (DNB) on the Suomi NPP satellite \cite{gravity_wave_data}. The VIIRS DNB captures broadband upwelling radiance in the visible spectrum, with a swath of approximately 3000 km and spatial resolution of about 1 km at nadir. Each 6-minute granule, typically measuring around 4000$\times$3000 pixels, is stored in Hierarchical Data Format version-5 (HDF5) and records radiance measurements in the 0.5--0.9\,$\mu$m wavelength range. These images capture extremely low radiance levels, on the order of $10^{-9}$\,W/$cm^{2}$\,sr$^{-1}$ \cite{murphy2006visible, gravitywavedata}, and face significant challenges due to various disturbances from city lights, clouds, and inherent instrumental noise, which are further compounded by limited availability of ground truth annotations.

To highlight the airglow associated with gravity wave events, nighttime images taken under new moon conditions are utilized, following established protocols described in \cite{miller2015upper, hu2019measuring}. The preprocessing pipeline involves several critical steps: first, subtracting the minimum pixel value from all pixels to establish a baseline, then scaling by the median value to normalize brightness variations, followed by normalizing the data to 0.5 to center the distribution, and finally transforming the intensity distribution from an approximate normal distribution to a uniform distribution while maintaining the original value range \cite{miller2013illuminating, mostafa2025gwavenet}. This preprocessing approach is essential for enhancing the visibility of subtle gravity wave patterns while reducing the impact of various noise sources. Figure~\ref{fig:vis-res} contains representative sample images that have been carefully annotated by domain experts, illustrating the complexity and variability of gravity wave patterns in the challenging VIIRS data environment.

\subsection{YOLO (You Only Look Once)}
YOLOv5 \cite{glenn_jocher_2020_4154370} is a state-of-the-art object detection model that integrates feature extraction, localization, and classification into a unified framework. We use YOLOv5-small version 6 due to dataset constraints and computational considerations. The architecture consists of three main components: the backbone (modified CSPDarknet) employs Conv, C3 (CSP bottleneck blocks for efficient feature learning), and SPPF (fast spatial pyramid pooling for multi-scale features) layers for feature extraction; the neck (PANet) aggregates features through bidirectional pathways for multi-scale information fusion; and the detection head predicts bounding boxes, objectness scores, and class probabilities at multiple scales.

YOLOv5 incorporates several optimization techniques including mosaic augmentation, auto-anchor optimization, cosine learning rate scheduling, and mixed-precision training. Its loss function combines binary cross-entropy for classification/objectness with CIoU loss for bounding box regression. The modular architecture of YOLOv5, particularly its convolutional operations in the backbone and neck components, makes it an excellent foundation for incorporating attention mechanisms and dilated convolutions. This flexibility allows for seamless integration of our proposed MDRC and SSCA components while maintaining the original detection capabilities, making YOLOv5 an ideal choice for vision model modifications and enhancements in specialized applications like gravity wave detection.

\section{Related Works}
\label{sec:relatedwork}

\subsection{Dilated Convolutions}
Dilated convolutions have proven to be highly effective in expanding receptive fields while maintaining spatial resolution across various computer vision applications. Zhou et al. demonstrated that dilated convolutions are capable of expanding the receptive field of feature points without sacrificing the resolution of the feature maps, which is particularly beneficial for semantic segmentation tasks \cite{zhou2018d}. Building upon this foundation, Liu et al. combined dilated convolutions with residual learning to improve road area extraction in semantic segmentation tasks, showing that the integration of these techniques enhances feature representation capabilities \cite{liu2019d}. Similarly, Zhang et al. applied dilated convolutions in their work, using atrous CNNs to capture a greater amount of semantic information for ultrasound image segmentation, demonstrating the versatility of dilated convolutions across different medical imaging domains \cite{zhang2020multiple}.

Furthermore, Chen et al. investigated the determination of effective dilation rates to aggregate multiscale features, thereby extending the receptive field while maintaining computational efficiency \cite{chen2021effective}. Their work provides important insights into optimal dilation rate selection, which influences our MDRC design. Additionally, Chen et al. showed that integrating dilated convolutions with spatial pyramid pooling can maintain a large receptive field while controlling the resolution of feature responses, leading to robust segmentation of objects at multiple scales \cite{chen2017deeplab}. Li et al. advanced this area by proposing a self-smoothing atrous convolution that naturally enhances the effectiveness of atrous convolutions for achieving larger receptive fields while reducing computational overhead \cite{li2021cascaded}. Wang et al. introduced smoothing techniques in dilated convolutions to alleviate gridding artifacts in dense predictions, addressing a common limitation of traditional dilated convolution approaches \cite{wang2018smoothed}.

\subsection{Attention Mechanisms}
Attention mechanisms have similarly been employed to boost performance in object detection and semantic segmentation tasks by enabling models to focus on the most relevant features and spatial regions. Park et al. introduced a pyramid attention mechanism that preserves semantic information in high-level features, which in turn enhances detection performance in feature pyramid networks by maintaining contextual information across different scales \cite{park2022pyramid}. Additionally, Zhou et al. proposed the Scale-aware Spatial Pyramid Pooling (SSPP) module together with Encoder Mask and Scale-Attention modules to address fundamental challenges such as scale-awareness, boundary sharpness, and long-range dependency modeling in semantic segmentation tasks \cite{zhou2020scale}. 

Cao et al. further contributed to this field by designing a network that integrates a Context Extraction Module with an Attention-guided Module, thereby enhancing object localization and recognition through the use of extensive contextual information and adaptive attention mechanisms \cite{cao2020attention}. Their approach demonstrates the effectiveness of combining contextual understanding with attention-based feature refinement. The influential work by Vaswani et al. introduced the transformer architecture with self-attention mechanisms \cite{vaswani2017attention}, which has been extensively adapted for computer vision tasks, providing the theoretical foundation for many subsequent attention-based approaches in visual recognition.

\subsection{Dilated Convolutions with Attention Mechanisms}
The combination of dilated convolutions with attention mechanisms has emerged as a promising approach to improve spatial information pooling and feature representation in computer vision tasks. Qiu et al. introduced the Attentive Atrous Spatial Pyramid Pooling (A2SPP) method, which merges Channel-Embedding Spatial Attention (CESA) with Spatial-Embedding Channel Attention (SECA) to effectively adapt to different feature scales, demonstrating superior performance in semantic segmentation applications \cite{qiu2022a2sppnet}. Similarly, Wang et al. proposed a salient object detection network that utilizes multi-scale saliency attention to focus on prominent regions by leveraging multi-scale saliency information, thereby achieving state-of-the-art performance with fast inference speeds \cite{wang2019salient}. Their work illustrates the synergistic benefits of combining multi-scale feature extraction with attention mechanisms.

\subsection{How Our Proposed Approach is different?}
Unlike existing approaches, our proposed system tackles the unique challenges of detecting gravity waves in VIIRS night band data by introducing several key innovations specifically designed for this challenging domain. Prior works typically apply dilated convolutions for general image processing tasks \cite{zhou2018d, liu2019d, chen2021effective}; however, our Multi Dilated Residual Convolution (MDRC) is specifically engineered to manage the significant variability in the scale, shape, and extent of gravity wave patterns, which exhibit characteristics fundamentally different from typical computer vision objects. Moreover, our Simplified Spatial and Channel Attention (SSCA) module provides a computationally efficient solution by integrating spatial and channel attention into a single unified operation, specifically optimized for detecting wave patterns in single-channel satellite imagery.

Our approach effectively handles significant variability in scale, shape, and pattern characteristics, even under high levels of interference from city lights and clouds, which is a defining characteristic of VIIRS data. These challenging conditions often hinder the performance of existing attention mechanisms \cite{park2022pyramid, zhou2020scale, cao2020attention}, whereas our approach demonstrates superior adaptability and robustness in such environments. Furthermore, previous combinations of dilated convolutions and attention techniques \cite{qiu2022a2sppnet, wang2019salient} were not designed to address the multi-scale nature of atmospheric gravity waves or the specific challenges of processing single-channel night band images with extremely low radiance levels, representing a significant gap that our proposed YOLOv5-based architecture effectively addresses.

\section{Methodology}
\label{sec:methodology}
\begin{figure}[h]
    \centering
    \includegraphics[width=.7\linewidth]{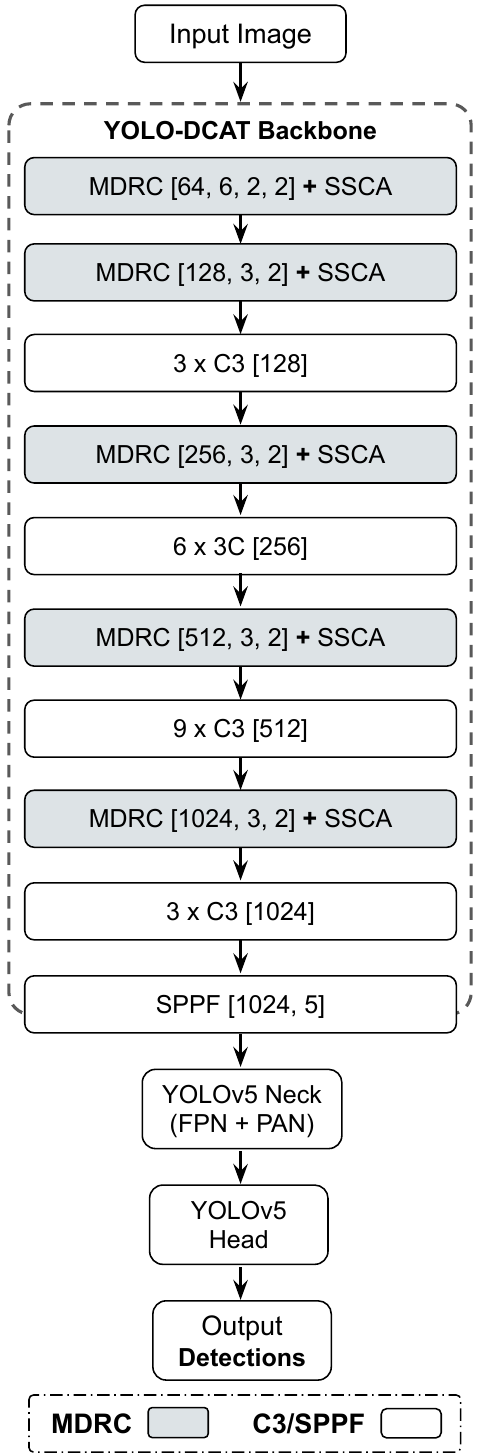}
    \caption{YOLO-DCAT architecture overview. The modified backbone, replacing the Convolution layers with MDRC+SSCA.}
    \label{fig:dcat-arch}
\end{figure}

The detection of gravity waves in complex satellite data demands specialized network modifications to overcome significant challenges, including scale variability, shape diversity, unwanted interference from city lights and clouds, and extremely low radiance values characteristic of VIIRS night band imagery. To address these challenges comprehensively, we enhance the YOLO architecture with two novel components: the Multi Dilated Residual Convolution (MDRC) and the Simplified Spatial and Channel Attention (SSCA) mechanism. These components work synergistically to improve feature extraction capabilities while maintaining computational efficiency, as illustrated in Figure \ref{fig:dcat-arch}.

\subsection{Multi Dilated Residual Convolution (MDRC)}
Standard convolution operations are inherently limited by a fixed receptive field, which significantly hinders the detection of gravity waves that exhibit variable scales and complex spatial patterns. Traditional convolutional layers with kernel size $k$ and stride $s$ can only capture features within a limited spatial context, making it challenging to detect objects that span different scales simultaneously.

\begin{figure*}[htb!]
    \centering
    \includegraphics[width=.75\linewidth]{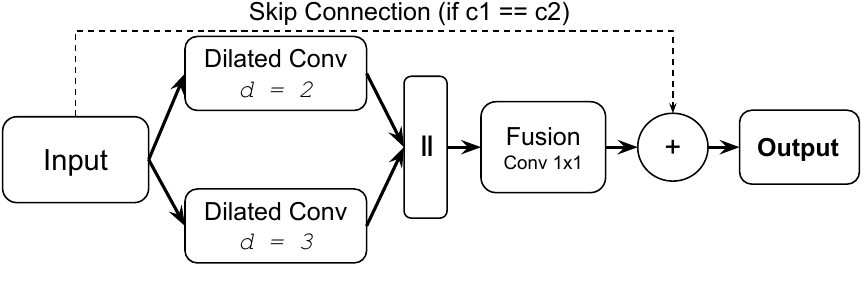}
    \caption{Multi Dilated Residual Convolution (MDRC) architecture.}
    \label{fig:mdrc}
\end{figure*}

Our Multi Dilated Residual Convolution (MDRC) addresses this fundamental limitation by incorporating multiple parallel dilated convolutions with different dilation rates, as shown in Figure~\ref{fig:mdrc}. Specifically, we utilize dilation rates of $d_1 = 2$ and $d_2 = 3$, which allow the model to capture features at different scales simultaneously while maintaining the original spatial resolution. The choice of these specific dilation rates is based on our empirical analysis presented in the ablation studies, where we systematically evaluated various combinations to determine the optimal configuration for gravity wave detection.

The mathematical formulation of our MDRC can be expressed as follows. For an input feature map $x \in \mathbb{R}^{H \times W \times C}$, where $H$, $W$, and $C$ represent height, width, and channel dimensions respectively, we apply parallel dilated convolutions:

\begin{equation}
\label{eq:dilated_conv}
y_i = x \ast W_i^{(d_i)}, \quad i \in \{1, 2\},
\end{equation}

\noindent where $\ast$ denotes the convolution operation, $W_i^{(d_i)}$ represents the weight parameters for the $i$-th dilated convolution with dilation rate $d_i$, and $y_i$ is the corresponding output feature map. The dilated convolution operation effectively increases the receptive field without increasing the number of parameters or computational cost significantly.

The outputs from both dilated convolutions are then concatenated along the channel dimension to combine multi-scale information:

\begin{equation}
\label{eq:concatenation}
F_{concat} = \text{Concat}(y_1, y_2),
\end{equation}

\noindent where $\text{Concat}(\cdot)$ represents the concatenation operation along the channel dimension, and $F_{concat} \in \mathbb{R}^{H \times W \times 2C}$ contains the combined multi-scale features.

To reduce the channel dimension back to the original size and incorporate the essential residual learning mechanism, we apply a $1 \times 1$ convolution followed by batch normalization and ReLU activation:

\begin{equation}
\label{eq:channel_reduction}
F_{reduced} = \text{ReLU}(\text{BN}(F_{concat} \ast W_{1 \times 1})),
\end{equation}

\noindent where $W_{1 \times 1}$ represents the weights of the $1 \times 1$ convolution layer, $\text{BN}(\cdot)$ denotes batch normalization, and $\text{ReLU}(\cdot)$ is the rectified linear unit activation function.

Finally, the output of the MDRC module is computed using the residual connection:

\begin{equation}
\label{eq:mdrc_output}
y_{MDRC} = x + F_{reduced},
\end{equation}

\noindent where the addition operation in Equation \ref{eq:mdrc_output} implements the skip connection that facilitates improved gradient flow during training, enabling the network to learn complex, multi-scale features more effectively while reducing the overall computational cost. This residual formulation allows the network to learn residual functions $F_{reduced} = F(x, \{W_i\})$, where $F(x, \{W_i\})$ represents the transformation learned by the dilated convolutions, following the principles established by He et al. \cite{he2016deep}.

The MDRC design effectively addresses the scale variability challenge in gravity wave detection by simultaneously capturing both fine-grained local patterns and broader spatial contexts, which is crucial for accurately detecting gravity waves that can appear at different scales within the same image.

\begin{figure*}[htb!]
    \centering
    \includegraphics[width=.75\textwidth]{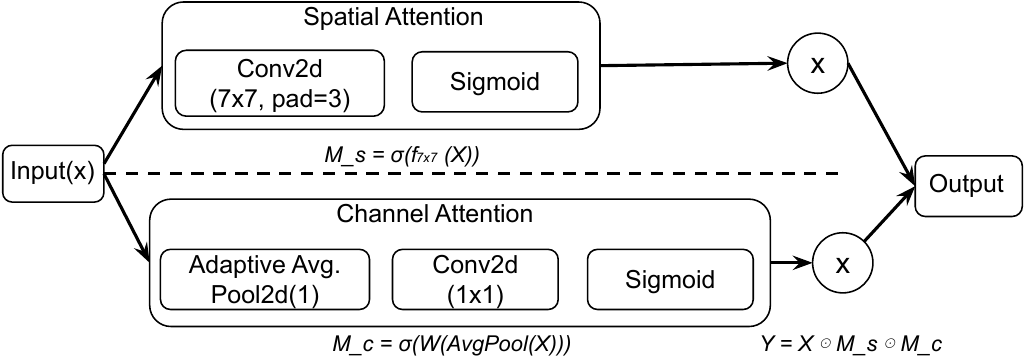}
    \caption{Simplified Spatial and Channel Attention (SSCA) architecture.}
    \label{fig:ssca}
\end{figure*}

\subsection{Simplified Spatial and Channel Attention (SSCA)}
Attention mechanisms, as originally introduced by Vaswani et al. \cite{vaswani2017attention}, enable models to focus on the most relevant parts of the input by computing relevance scores. In the context of computer vision, attention mechanisms help models selectively focus on important spatial locations and channel features while suppressing irrelevant information.

The fundamental principle of attention can be expressed through the computation of attention weights. For a given query $q_i$ and key $k_j$, the attention energy is calculated as:

\begin{equation}
\label{eq:attention_energy}
e_{ij} = q_i \cdot k_j,
\end{equation}

These energy scores are then normalized using the softmax function to obtain attention weights:

\begin{equation}
\label{eq:attention_weights}
\alpha_{ij} = \frac{\exp(e_{ij})}{\sum_{k=1}^{N} \exp(e_{ik})},
\end{equation}

\noindent where $N$ represents the total number of keys, and $\alpha_{ij}$ represents the normalized attention weight indicating the relevance of key $k_j$ to query $q_i$.

Building upon these fundamental principles and inspired by the Convolutional Block Attention Module (CBAM) \cite{woo2018cbam}, we propose the Simplified Spatial and Channel Attention (SSCA) mechanism to further refine feature extraction for gravity wave detection. CBAM applies channel and spatial attention sequentially, where channel attention is computed as:

\begin{equation}
\label{eq:cbam_channel}
M_c(X) = \sigma\left(\text{MLP}(\text{AvgPool}(X)) + \text{MLP}(\text{MaxPool}(X))\right),
\end{equation}

and spatial attention is computed as:

\begin{equation}
\label{eq:cbam_spatial}
M_s(X) = \sigma\left(f^{7\times7}\left([\text{AvgPool}_c(X); \text{MaxPool}_c(X)]\right)\right),
\end{equation}

\noindent where $\sigma(\cdot)$ represents the sigmoid activation function, $\text{MLP}(\cdot)$ denotes a multi-layer perceptron, and $f^{7\times7}(\cdot)$ represents a $7 \times 7$ convolution operation.

However, the sequential application of attention mechanisms in CBAM can be computationally expensive and may not be optimal for detecting gravity wave patterns that require simultaneous consideration of both spatial and channel information. Therefore, we propose SSCA, which computes spatial and channel attention simultaneously in a single, efficient operation, as illustrated in Figure \ref{fig:ssca}.

For an input feature map $X \in \mathbb{R}^{H \times W \times C}$, our SSCA mechanism first computes spatial attention by aggregating channel information through both average and max pooling operations:

\begin{equation}
\label{eq:ssca_spatial_pooling}
X_{avg} = \text{AvgPool}_c(X), \quad X_{max} = \text{MaxPool}_c(X),
\end{equation}

\noindent where $\text{AvgPool}_c(\cdot)$ and $\text{MaxPool}_c(\cdot)$ represent average and max pooling operations along the channel dimension, respectively, resulting in feature maps $X_{avg}, X_{max} \in \mathbb{R}^{H \times W \times 1}$.

The spatial attention map is then computed by concatenating these pooled features and applying a $7 \times 7$ convolution followed by sigmoid activation:

\begin{equation}
\label{eq:ssca_spatial}
M_s = \sigma\left(f^{7\times7}_{avg}\left([X_{max}; X_{avg}]\right)\right),
\end{equation}

\noindent where $[X_{max}; X_{avg}]$ denotes concatenation along the channel dimension, resulting in a tensor of size $H \times W \times 2$, and $f^{7\times7}_{avg}(\cdot)$ represents a $7 \times 7$ convolution with a single output channel.

Simultaneously, the channel attention is computed by applying global average pooling followed by a compact fully connected layer sequence:

\begin{equation}
\label{eq:ssca_channel}
M_c = \sigma\left(W_2\left(\text{ReLU}(W_1(\text{AvgPool}(X)))\right)\right),
\end{equation}

\noindent where $\text{AvgPool}(X) \in \mathbb{R}^{1 \times 1 \times C}$ represents global average pooling that compresses spatial dimensions, $W_1 \in \mathbb{R}^{C/r \times C}$ and $W_2 \in \mathbb{R}^{C \times C/r}$ are the weights of the fully connected layers with reduction ratio $r$ (typically set to 16), and $\text{ReLU}(\cdot)$ is the rectified linear unit activation function.

The final output of the SSCA mechanism is obtained by applying both attention maps element-wise to the input feature map:

\begin{equation}
\label{eq:ssca_output}
Y = X \odot M_s \odot M_c,
\end{equation}

\noindent where $\odot$ denotes element-wise multiplication, and the broadcasting mechanism automatically handles the dimension compatibility between $X \in \mathbb{R}^{H \times W \times C}$, $M_s \in \mathbb{R}^{H \times W \times 1}$, and $M_c \in \mathbb{R}^{1 \times 1 \times C}$.

This combined mechanism, as expressed in Equation \ref{eq:ssca_output}, directs the network's focus to the most salient spatial regions and channels simultaneously, which is crucial for effectively detecting gravity waves in the challenging VIIRS night band data environment. The simultaneous computation of both attention types reduces computational overhead compared to sequential approaches while maintaining the benefits of both spatial and channel attention.

The integration of MDRC and SSCA into the YOLOv5 framework, as shown in the overall architecture in Figure \ref{fig:dcat-arch}, enhances the model's capacity to manage multi-scale features through the dilated convolution mechanism described in Equations \ref{eq:dilated_conv}-\ref{eq:mdrc_output}, while improving its focus on relevant information through the attention mechanism detailed in Equations \ref{eq:ssca_spatial_pooling}-\ref{eq:ssca_output}. This synergistic combination results in more accurate gravity wave localization from challenging VIIRS night band data, as demonstrated in our comprehensive experimental evaluation.

\section{Experiments and Results}
\label{sec:experiments}

In this section, we systematically compare our proposed modifications against the baseline YOLO model to evaluate their individual and combined impact on detection performance. Subsequently, we extend our analysis to compare with state-of-the-art attention mechanisms and examine their integration with multi-scale feature extraction techniques. Our experimental evaluation is designed to demonstrate the effectiveness of each component and validate the overall approach for gravity wave detection in challenging satellite imagery.

\subsection{Experimental Setup}
For accurate detection and evaluation, we employed the \texttt{LabelImg} software to annotate objects of interest with precise bounding boxes defined by their top-left $(x_1, y_1)$ and bottom-right $(x_2, y_2)$ coordinates. These detailed annotations enable YOLO to compute Intersection over Union (IoU) scores by evaluating the spatial overlap between predicted and ground truth bounding boxes using the standard IoU formula:

\begin{equation}
\text{IoU} = \frac{\text{Area of Overlap}}{\text{Area of Union}} = \frac{|B_p \cap B_{gt}|}{|B_p \cup B_{gt}|},
\end{equation}

\noindent where $B_p$ represents the predicted bounding box and $B_{gt}$ represents the ground truth bounding box.

We defined two distinct classes for our detection task: `gw' for the gravity wave regions of interest, and `noise' for unwanted interference events such as city lights and clouds in the satellite imagery. Our comprehensive dataset comprises 600 carefully annotated gravity wave instances, and all models were systematically trained, validated, and tested using a standard 70:20:10 split to ensure fair comparison and reliable evaluation metrics.

We integrated our proposed methods, including the Multi Dilated Residual Convolution (MDRC) detailed in Equations \ref{eq:dilated_conv}-\ref{eq:mdrc_output} and the Simplified Spatial and Channel Attention (SSCA) mechanism described in Equations \ref{eq:ssca_spatial_pooling}-\ref{eq:ssca_output}, as well as state-of-the-art comparative techniques (Transformer, ViT, CBAM) into the YOLOv5 framework as independent modular functions. All implementations maintain consistent training parameters and optimization settings to ensure fair experimental comparison.


\subsection{Results and Analysis}

\begin{table*}[ht]
  \caption{Comprehensive comparison of the YOLO baseline model with our proposed method, systematically evaluating MDRC and SSCA individually, as well as their combined approach $\text{Y}_{\text{base}} + \text{MDRC} + \text{SSCA}$.}
  \label{tab:baseline-proposed}
  \centering \large
  \begin{tabular}{l|c|c|c|c|c} 
    \hline
    Method & Precision (\%) & Recall (\%) & mAP50 (\%) & mAP50-95 (\%) & IoU (\%)\\ 
    \hline
    Y\textsubscript{base} & 56.20 & 39.70 & 41.80 & 16.30 & 31.62\\
    Y\textsubscript{base}$+$MDRC        & 57.00 & 49.20 & 49.90 & 24.10 & 32.76\\
    Y\textsubscript{base}$+$SSCA        & 57.60 & 56.10 & 50.90 & 24.80 & 38.64\\
    Y\textsubscript{base}$+$MDRC$+$SSCA & \textbf{58.80} & \textbf{66.70} & \textbf{55.30} & \textbf{26.60} & \textbf{48.74}\\
    \hline
  \end{tabular}
\end{table*}

Table~\ref{tab:baseline-proposed} presents a detailed performance comparison between the YOLO baseline model and our proposed modifications, demonstrating the progressive improvement achieved by each component. The YOLO baseline ($\text{Y}_{\text{base}}$) achieves a mean average precision (mAP50) of 41.80\% and an intersection-over-union (IoU) of 31.62\%, establishing the performance baseline for gravity wave detection. 

The incorporation of MDRC alone significantly improves recall from 39.70\% to 49.20\% and mAP50 to 49.90\%, representing an 8.1 percentage point improvement in mAP50. This substantial improvement highlights the effectiveness of multi-dilated convolutions in capturing multi-scale gravity wave patterns through the parallel processing approach detailed in Equations \ref{eq:dilated_conv}-\ref{eq:mdrc_output}. The MDRC's ability to simultaneously capture features at different scales (dilation rates of 2 and 3) proves crucial for detecting gravity waves that exhibit significant scale variability.

The addition of SSCA alone further improves performance substantially, with mAP50 reaching 50.90\% (a 9.1 percentage point improvement over baseline) and IoU enhancing to 38.64\% (a 7.02 percentage point improvement). This demonstrates the significant benefit of the simultaneous spatial and channel attention mechanism described in Equations \ref{eq:ssca_spatial_pooling}-\ref{eq:ssca_output} in refining feature representation and focusing on relevant gravity wave patterns while suppressing background interference.

The optimal performance is achieved when both MDRC and SSCA are combined, yielding the highest recall of 66.70\% (27 percentage points above baseline), mAP50 of 55.30\% (13.5 percentage points above baseline), and IoU of 48.74\% (17.12 percentage points above baseline). This confirms that integrating both mechanisms creates a synergistic effect that significantly enhances the model's ability to detect gravity waves in challenging VIIRS data through improved multi-scale feature extraction and attention-guided processing.

\begin{table*}[ht]
  \caption{Systematic comparison of the YOLO baseline with state-of-the-art attention-based models and our proposed SSCA method, demonstrating the effectiveness of our attention mechanism design.}
  \label{tab:sota-att}
  \centering \large
  \begin{tabular}{l|c|c|c|c|c} 
    \hline
    Method & Precision (\%) & Recall (\%) & mAP50 (\%) & mAP50-95 (\%) & IoU (\%)\\ 
    \hline
    Y\textsubscript{base}$+$Transformer & 44.70 & 46.10 & 43.20 & 13.60 & 34.48\\
    Y\textsubscript{base}$+$ViT         & 46.40 & 49.50 & 46.40 & 19.90 & 34.66\\
    Y\textsubscript{base}$+$CBAM        & 51.80 & \textbf{56.70} & 50.40 & 21.70 & 36.92\\
    Y\textsubscript{base}$+$SSCA (ours) & \textbf{57.60} & 56.10 & \textbf{50.90} & \textbf{24.80} & \textbf{38.64}\\ 
    \hline
  \end{tabular}
\end{table*}

Table~\ref{tab:sota-att} extends the comparison by evaluating our SSCA method against state-of-the-art attention-based models, providing insights into the effectiveness of different attention mechanisms for gravity wave detection. Traditional transformer-based attention mechanisms, including ViT and standard Transformer models, struggle significantly with the challenging single-channel VIIRS data, achieving lower mAP50 scores of 46.40\% and 43.20\%, respectively. These transformer-based approaches, while successful in natural image processing, face difficulties in handling the specific characteristics of gravity wave patterns in satellite imagery, including their subtle appearance and interference from various noise sources.

The Convolutional Block Attention Module (CBAM), which sequentially applies spatial and channel attention as described in Equations \ref{eq:cbam_channel} and \ref{eq:cbam_spatial}, shows notable improvement with a mAP50 of 50.40\% and IoU of 36.92\%. CBAM's performance demonstrates the value of attention mechanisms in gravity wave detection, but the sequential processing approach limits its effectiveness compared to our simultaneous attention computation.

Our SSCA method significantly outperforms all other attention-based models, achieving the highest mAP50 of 50.90\% (4.5 percentage points above CBAM) and IoU of 38.64\% (1.72 percentage points above CBAM). This superior performance confirms that SSCA, by computing both spatial and channel attention simultaneously through the mechanism detailed in Equations \ref{eq:ssca_spatial_pooling}-\ref{eq:ssca_output}, provides a more efficient and effective solution for gravity wave detection. The simultaneous computation not only reduces computational overhead but also enables better integration of spatial and channel information, which is crucial for detecting subtle gravity wave patterns.

\begin{table*}[ht]
  \caption{Comprehensive comparison of the YOLO baseline with state-of-the-art attention-based models incorporating the MDRC module and our proposed full YOLO-DCAT method. (\textit{YOLO-DCAT = $\text{Y}_{\text{base}} + \text{MDRC} + \text{SSCA}$}).}
  \label{tab:sota-full}
  \centering \large
  \begin{tabular}{l|c|c|c|c|c} 
    \hline
    Method & Precision (\%) & Recall (\%) & mAP50 (\%) & mAP50-95 (\%) & IoU (\%)\\ 
    \hline
    Y\textsubscript{base}$+$MDRC$+$Transformer  & 49.40 & 50.40 & 44.50 & 17.90 & 37.71\\
    Y\textsubscript{base}$+$MDRC$+$ViT          & 51.90 & 58.80 & 49.20 & 21.30 & 38.12\\
    Y\textsubscript{base}$+$MDRC$+$CBAM         & \textbf{59.50} & 59.10 & 52.80 & 23.80 & 44.48\\
    YOLO-DCAT (ours)  & 58.80 & \textbf{66.70} & \textbf{55.30} & \textbf{26.60} & \textbf{48.74}\\
    \hline
  \end{tabular}
\end{table*}

Table~\ref{tab:sota-full} presents a comprehensive comparison of models that incorporate both MDRC and various attention mechanisms, demonstrating the synergistic benefits of combining multi-scale feature extraction with attention mechanisms. The inclusion of MDRC consistently improves performance across all tested attention mechanisms, with CBAM reaching a mAP50 of 52.80\% and IoU of 44.48\% when combined with MDRC. This improvement demonstrates that the multi-scale feature extraction capability of MDRC, as formulated in Equations \ref{eq:dilated_conv}-\ref{eq:mdrc_output}, enhances the effectiveness of all attention mechanisms by providing richer feature representations.

However, our proposed full model ($\text{Y}_{\text{base}} + \text{MDRC} + \text{SSCA}$) achieves superior results across most metrics, with the highest mAP50 of 55.30\% (2.5\% above MDRC+CBAM), recall of 66.70\% (7.6\% above MDRC+CBAM), and IoU of 48.74\% (4.26\% above MDRC+CBAM). These results highlight the importance of the integrated design approach, where MDRC provides multi-scale feature extraction capabilities while SSCA enables efficient simultaneous attention computation. The combination outperforms Transformer, ViT, and CBAM models, even when each is enhanced with MDRC, demonstrating that our approach effectively addresses the specific challenges of gravity wave detection in satellite imagery.

\begin{table}[htb!]
  \caption{Statistical analysis showing mean and standard deviation comparison of $\text{Y}_{\text{base}}$ with the proposed MDRC, SSCA, and all other state-of-the-art models, demonstrating robustness and consistency. (\textit{YOLO-DCAT = $\text{Y}_{\text{base}} + \text{MDRC} + \text{SSCA}$}).}
  \label{table:mean-std-comparison}
  \centering \large
  \begin{tabular}{lcc}
    \toprule
    Model & mAP50 (\%) & IoU (\%)\\
    \midrule
    Y\textsubscript{base}       
                                & 37.80$\pm$4.30 & 28.74$\pm$3.80\\
    Y\textsubscript{base}$+$MDRC (ours)
                                & 44.20$\pm$4.10 & 28.94$\pm$3.80\\
    Y\textsubscript{base}$+$SSCA (ours)
                                & 48.40$\pm$2.50 & 36.86$\pm$2.10\\ 
    Y\textsubscript{base}$+$MDRC$+$Transformer  
                                & 41.60$\pm$4.80 & 35.71$\pm$3.50 \\
    Y\textsubscript{base}$+$MDRC$+$ViT
                                & 46.10$\pm$3.30 & 36.42$\pm$2.60 \\
    Y\textsubscript{base}$+$MDRC$+$CBAM
                                & 50.90$\pm$2.40 & 42.68$\pm$2.20 \\
    YOLO-DCAT (ours) 
                                & \textbf{54.50$\pm$1.20} & \textbf{46.78$\pm$1.90} \\
    \bottomrule
  \end{tabular}
\end{table}

Table~\ref{table:mean-std-comparison} presents a crucial statistical comparison of the mean and standard deviation of mAP50 and IoU metrics across different models, providing insights into both performance and stability. The YOLO baseline model exhibits the lowest mAP50 of 37.80\% and IoU of 28.74\%, with substantial variability (standard deviations of 4.30\% and 3.80\% respectively), indicating inconsistent performance across different runs and data splits.

Adding MDRC alone improves mAP50 to 44.20\% while maintaining similar variability, though its IoU improvement is minimal (28.94\%). In contrast, SSCA alone shows greater improvement with mAP50 reaching 48.40\% and IoU increasing to 36.86\%, accompanied by reduced standard deviations (2.50\% and 2.10\% respectively), indicating improved stability and consistency.

Among other attention-based models, CBAM achieves a competitive mAP50 of 50.90\% with reasonable stability (standard deviation of 2.40\%), while transformer-based approaches display higher variability and lower performance. Notably, our full model ($\text{Y}_{\text{base}} + \text{MDRC} + \text{SSCA}$) achieves both the highest performance and the most stable results, with mAP50 reaching 54.50\% and IoU of 46.78\%, accompanied by the lowest standard deviations (1.20\% and 1.90\% respectively). This combination of high performance and low variability confirms the robustness and reliability of our integrated approach in detecting gravity waves under challenging and variable conditions.

\subsection{Comprehensive Ablation Study}
We conducted extensive ablation studies to systematically analyze the impact of different design choices and validate the effectiveness of our proposed components. These studies provide crucial insights into optimal configurations and demonstrate the importance of proper component placement within the network architecture.

\begin{table}
  \caption{Systematic comparison of applying dilation rates ($d=2,3$) in different layers of YOLOv5, revealing the critical importance of proper placement for optimal performance.}
  \label{tab:layer_dilation}
  \centering \large
  \begin{tabular}{l|c|c} 
    \hline
    Dilation Effects & mAP50(\%) & IoU(\%)\\
    \hline
    Y\textsubscript{base} (No dilation) & 41.80 & 31.62 \\
    \hline
    MDRC in C3 layers ($d=2,3$) & 34.20 & 27.75 \\
    MDRC in Conv layers ($d=2,3$) & \textbf{49.90} & \textbf{38.92} \\
    \hline
  \end{tabular}
\end{table}

Table~\ref{tab:layer_dilation} reveals the critical importance of where dilation is applied within the YOLOv5 architecture, providing essential insights for optimal implementation. Interestingly, implementing MDRC in C3 layers actually degrades performance significantly below the baseline YOLOv5 (34.20\% vs. 41.80\% mAP50, representing a 7.6 percentage point decrease). This substantial performance degradation likely occurs because C3 layers employ cross-stage partial connections that fundamentally alter the feature flow patterns. When dilated convolutions are applied within these layers, they may disrupt the carefully designed cross-stage connections, leading to suboptimal feature representations that cannot effectively capture the complex shapes and patterns characteristic of gravity waves with varying scales.

In stark contrast, our empirical evaluations demonstrate that multi-dilated convolutions work significantly better when applied to standard convolutional layers, achieving a mAP50 of 49.90\% and IoU of 38.92\%, representing improvements of 8.1 and 7.3 percentage points respectively over the baseline. This substantial improvement validates our architectural design choice and demonstrates that proper placement of dilated convolutions is crucial for maintaining the benefits of multi-scale feature extraction while preserving the original network's feature processing capabilities.

\begin{table}
  \caption{Comprehensive analysis of various dilation rates and their combinations on gravity wave detection performance, demonstrating optimal configuration selection.}
  \label{tab:dial}
  \centering \large
  \begin{tabular}{l|c|c} 
    \hline
    Approach & mAP50(\%) & IoU(\%)\\
    \hline
    Y\textsubscript{base}$+$MDRC ($d=2$) & 45.80 & 36.42 \\
    Y\textsubscript{base}$+$MDRC ($d=3$) & 44.60 & 35.87 \\
    Y\textsubscript{base}$+$MDRC ($d=4$) & 39.90 & 29.54 \\
    \hline
    Y\textsubscript{base}$+$MDRC ($d=2,3$) & \textbf{49.90} & \textbf{38.92} \\
    Y\textsubscript{base}$+$MDRC ($d=2,4$) & 42.30 & 31.81 \\
    Y\textsubscript{base}$+$MDRC ($d=3,4$) & 41.50 & 31.15 \\
    \hline
  \end{tabular}
\end{table}

Table~\ref{tab:dial} provides a systematic comparison of various dilation rates and their combinations, offering crucial insights into optimal parameter selection for gravity wave detection. Single dilation rates show moderate improvements over the baseline, with $d=2$ achieving 45.80\% mAP50 and $d=3$ reaching 44.60\% mAP50. However, $d=4$ performs significantly worse (39.90\% mAP50), even falling below the baseline performance, suggesting that excessively large dilation rates capture features at scales that are less relevant or potentially harmful for gravity wave detection.

The combination results reveal important patterns in multi-scale feature extraction effectiveness. Our proposed combination of $d=2,3$ significantly outperforms all other options with mAP50 of 49.90\% and IoU of 38.92\%, representing the optimal balance between capturing fine-grained local patterns ($d=2$) and broader spatial contexts ($d=3$). Combinations involving $d=4$ consistently show diminished performance, with $d=2,4$ achieving only 42.30\% mAP50 and $d=3,4$ reaching 41.50\% mAP50. This systematic degradation when including larger dilation rates suggests that gravity waves in VIIRS data exhibit optimal detectable patterns within specific scale ranges, and that excessive dilation may introduce irrelevant contextual information or spatial artifacts that interfere with accurate detection.

\begin{table}
  \caption{Systematic evaluation of applying SSCA in different layers of YOLOv5, confirming the importance of proper attention mechanism placement for optimal performance.}
  \label{tab:ssca_layer}
  \centering \large
  \begin{tabular}{l|c|c} 
    \hline
    Attention Application & mAP50(\%) & IoU(\%)\\
    \hline
    Y\textsubscript{base} (No attention) & 41.80 & 31.62 \\
    \hline
    SSCA in C3 layers & 36.70 & 29.15 \\
    SSCA in Conv layers & \textbf{50.90} & \textbf{38.64} \\
    \hline
  \end{tabular}
\end{table}

Table~\ref{tab:ssca_layer} demonstrates the significant impact of SSCA placement within the YOLOv5 architecture, mirroring the patterns observed with MDRC placement. Implementing SSCA in C3 layers degrades performance substantially below the baseline (36.70\% vs. 41.80\% mAP50, representing a 5.1 percentage point decrease), while applying it to convolutional layers significantly improves results (50.90\% mAP50, representing a 9.1 percentage point improvement).

This consistent pattern aligns with our MDRC findings and suggests that C3 layers, with their existing cross-stage connections and bottleneck structures, may create conflicts when combined with additional attention mechanisms. The bottleneck design in C3 layers is specifically optimized for computational efficiency through channel reduction and expansion, and introducing attention mechanisms may disrupt this carefully balanced information flow. In contrast, convolutional layers, being earlier in the feature extraction pipeline, benefit significantly from attention guidance as they establish the initial feature representations without architectural interference. This observation confirms that both MDRC and SSCA achieve optimal performance when applied to standard convolutional layers, validating our integrated architectural design approach.

\section{Discussions}
\label{sec:disc}

The experimental results confirm the effectiveness of our proposed YOLO-DCAT architecture for gravity wave detection in challenging satellite imagery. The baseline YOLO model exhibits limited performance, underscoring the inherent difficulties in detecting subtle gravity wave patterns amid the challenging characteristics of VIIRS night band data, including extremely low radiance values, significant interference from city lights and clouds, and high variability in scale, shape, and spatial extent of gravity wave manifestations.

Our comprehensive ablation studies offer key insights into optimal architectural design principles. Table~\ref{tab:layer_dilation} shows that applying the Multi Dilated Residual Convolution (MDRC) in C3 layers degrades performance substantially, likely because the cross-stage partial connections in these layers fail to effectively capture the complex shapes and multi-scale characteristics of gravity waves. The cross convolutions within C3 layers may disrupt the carefully designed feature flow patterns needed for detecting variable-scale atmospheric phenomena. In contrast, using MDRC in standard convolutional layers significantly improves performance, demonstrating the importance of proper architectural placement for dilated convolution mechanisms.

Table~\ref{tab:dial} further indicates that while single dilation rates provide moderate performance gains, the combination of dilation rates $d=2,3$ yields optimal results by capturing both fine-grained local patterns and broader spatial contexts simultaneously. Larger dilation rates such as $d=4$ capture features at scales that are less relevant for gravity wave detection, potentially introducing spatial artifacts or irrelevant contextual information that interferes with accurate pattern recognition.

Integrating the optimized MDRC and SSCA components into the YOLOv5 framework yields a synergistic improvement across all evaluation metrics, confirming that the multi-scale feature extraction and simultaneous attention mechanisms complement each other effectively. Our approach significantly outperforms state-of-the-art attention mechanisms, such as ViT and Transformer, which struggled with the specific characteristics of single-channel satellite imagery, and substantially improves upon YOLO baseline models under challenging detection conditions. Although CBAM remained quite competitive, our SSCA mechanism demonstrates superior performance through its efficient simultaneous computation of spatial and channel attention.

Figure~\ref{fig:vis-res} illustrates the impressive localization performance of our approach compared to the baseline and other advanced models, providing visual evidence of the practical improvements achieved by our architectural enhancements. Our method not only detects gravity waves more accurately relative to the expert-labeled ground truth data, but it also assigns higher confidence scores than both the state-of-the-art and baseline models, which is crucial for automated atmospheric monitoring applications. In particular, the bottom row of Figure~\ref{fig:vis-res} demonstrates that while other methods struggle with correct localization or even mis-localize gravity wave patterns, our proposed approach consistently outperforms them, including the baseline models, even in challenging scenarios where gravity waves are partially obscured or embedded within clouds. In the top row, apart from the YOLO base model, the localization results of all other methods are closer to those of our proposed approach, but with notably lower confidence scores, indicating the superior reliability of our detection system.

\noindent \textit{In the future}, we plan to focus on developing more comprehensive attention mechanisms to extend our approach to a wider range of atmospheric and oceanic phenomena, including ocean eddies \cite{mostafa2023cnn}, mesospheric bores \cite{hozumi2019geographical}, cloud properties \cite{tushar2024cloudunet}, and mesoscale hurricanes \cite{hasan2025comparison}. We also plan to explore integration with advanced architectures such as YOLOv12 \cite{tian2025yolov12} and DETR \cite{carion2020end}, and to investigate additional feature extraction techniques, including residual attention for denoising and combined local-global attention for enhanced contextual understanding of complex atmospheric patterns.

\section{Conclusion}
\label{sec:conclusions}
Accurate localization of atmospheric gravity waves is crucial for understanding their global impact, yet satellite imagery presents significant challenges due to the high variability in wave patterns, interference from city lights and clouds, and the limitations of single-band datasets. To address these challenges, we propose YOLO-DCAT, an enhanced YOLOv5 model incorporating Multi Dilated Residual Convolution (MDRC) and Simplified Spatial and Channel Attention (SSCA), designed to improve gravity wave detection in complex satellite imagery. MDRC captures multi-scale features through parallel dilated convolutions with varying dilation rates, while SSCA focuses on relevant spatial regions and channel features to enhance detection accuracy while suppressing noise interference. Our approach effectively mitigates these difficulties and enhances localization accuracy, as demonstrated by our experimental results which show substantial improvements over baseline and state-of-the-art methods. These advancements establish a robust framework for analyzing challenging satellite data, thereby enhancing climate modeling, promoting sustainable development, and advancing atmospheric research through computer vision.


\bibliographystyle{ACM-Reference-Format}
\bibliography{main}


\begin{thebibliography}{35}


\ifx \showCODEN    \undefined \def \showCODEN     #1{\unskip}     \fi
\ifx \showISBNx    \undefined \def \showISBNx     #1{\unskip}     \fi
\ifx \showISBNxiii \undefined \def \showISBNxiii  #1{\unskip}     \fi
\ifx \showISSN     \undefined \def \showISSN      #1{\unskip}     \fi
\ifx \showLCCN     \undefined \def \showLCCN      #1{\unskip}     \fi
\ifx \shownote     \undefined \def \shownote      #1{#1}          \fi
\ifx \showarticletitle \undefined \def \showarticletitle #1{#1}   \fi
\ifx \showURL      \undefined \def \showURL       {\relax}        \fi
\providecommand\bibfield[2]{#2}
\providecommand\bibinfo[2]{#2}
\providecommand\natexlab[1]{#1}
\providecommand\showeprint[2][]{arXiv:#2}

\bibitem[Alexander and Holton(1997)]%
        {alexander1997model}
\bibfield{author}{\bibinfo{person}{MJ Alexander} {and} \bibinfo{person}{James~R Holton}.} \bibinfo{year}{1997}\natexlab{}.
\newblock \showarticletitle{A model study of zonal forcing in the equatorial stratosphere by convectively induced gravity waves}.
\newblock \bibinfo{journal}{\emph{Journal of the atmospheric sciences}} \bibinfo{volume}{54}, \bibinfo{number}{3} (\bibinfo{year}{1997}), \bibinfo{pages}{408--419}.
\newblock


\bibitem[Cao et~al\mbox{.}(2020)]%
        {cao2020attention}
\bibfield{author}{\bibinfo{person}{Junxu Cao}, \bibinfo{person}{Qi Chen}, \bibinfo{person}{Jun Guo}, {and} \bibinfo{person}{Ruichao Shi}.} \bibinfo{year}{2020}\natexlab{}.
\newblock \showarticletitle{Attention-guided context feature pyramid network for object detection}.
\newblock \bibinfo{journal}{\emph{arXiv preprint arXiv:2005.11475}} (\bibinfo{year}{2020}).
\newblock


\bibitem[Carion et~al\mbox{.}(2020)]%
        {carion2020end}
\bibfield{author}{\bibinfo{person}{Nicolas Carion}, \bibinfo{person}{Francisco Massa}, \bibinfo{person}{Gabriel Synnaeve}, \bibinfo{person}{Nicolas Usunier}, \bibinfo{person}{Alexander Kirillov}, {and} \bibinfo{person}{Sergey Zagoruyko}.} \bibinfo{year}{2020}\natexlab{}.
\newblock \showarticletitle{End-to-end object detection with transformers}. In \bibinfo{booktitle}{\emph{European conference on computer vision}}. Springer, \bibinfo{pages}{213--229}.
\newblock


\bibitem[Chen and Lin(2021)]%
        {chen2021effective}
\bibfield{author}{\bibinfo{person}{Hanshen Chen} {and} \bibinfo{person}{Huiping Lin}.} \bibinfo{year}{2021}\natexlab{}.
\newblock \showarticletitle{An effective hybrid atrous convolutional network for pixel-level crack detection}.
\newblock \bibinfo{journal}{\emph{IEEE Transactions on Instrumentation and Measurement}}  \bibinfo{volume}{70} (\bibinfo{year}{2021}), \bibinfo{pages}{1--12}.
\newblock


\bibitem[Chen et~al\mbox{.}(2017)]%
        {chen2017deeplab}
\bibfield{author}{\bibinfo{person}{Liang-Chieh Chen}, \bibinfo{person}{George Papandreou}, \bibinfo{person}{Iasonas Kokkinos}, \bibinfo{person}{Kevin Murphy}, {and} \bibinfo{person}{Alan~L Yuille}.} \bibinfo{year}{2017}\natexlab{}.
\newblock \showarticletitle{Deeplab: Semantic image segmentation with deep convolutional nets, atrous convolution, and fully connected crfs}.
\newblock \bibinfo{journal}{\emph{IEEE transactions on pattern analysis and machine intelligence}} \bibinfo{volume}{40}, \bibinfo{number}{4} (\bibinfo{year}{2017}), \bibinfo{pages}{834--848}.
\newblock


\bibitem[{Curtis Seaman}(2013)]%
        {gravity_wave_data}
\bibfield{author}{\bibinfo{person}{{Curtis Seaman}}.} \bibinfo{year}{2013}\natexlab{}.
\newblock \bibinfo{title}{{Beginning to See the Light: an Introduction to VIIRS DNB and NCC}}.
\newblock \bibinfo{howpublished}{\url{https://rammb.cira.colostate.edu/projects/alaska/blog/index.php/uncategorized/beginning-to-see-the-light-an-introduction-to-viirs-dnb-and-ncc/}, [online accessed: 2025-02-28]}.
\newblock


\bibitem[Dosovitskiy et~al\mbox{.}(2020)]%
        {dosovitskiy2020image}
\bibfield{author}{\bibinfo{person}{Alexey Dosovitskiy}, \bibinfo{person}{Lucas Beyer}, \bibinfo{person}{Alexander Kolesnikov}, \bibinfo{person}{Dirk Weissenborn}, \bibinfo{person}{Xiaohua Zhai}, \bibinfo{person}{Thomas Unterthiner}, \bibinfo{person}{Mostafa Dehghani}, \bibinfo{person}{Matthias Minderer}, \bibinfo{person}{Georg Heigold}, \bibinfo{person}{Sylvain Gelly}, {et~al\mbox{.}}} \bibinfo{year}{2020}\natexlab{}.
\newblock \showarticletitle{An image is worth 16x16 words: Transformers for image recognition at scale}.
\newblock \bibinfo{journal}{\emph{arXiv preprint arXiv:2010.11929}} (\bibinfo{year}{2020}).
\newblock


\bibitem[Fritts and Alexander(2003)]%
        {fritts2003gravity}
\bibfield{author}{\bibinfo{person}{David~C Fritts} {and} \bibinfo{person}{M~Joan Alexander}.} \bibinfo{year}{2003}\natexlab{}.
\newblock \showarticletitle{Gravity wave dynamics and effects in the middle atmosphere}.
\newblock \bibinfo{journal}{\emph{Reviews of geophysics}} \bibinfo{volume}{41}, \bibinfo{number}{1} (\bibinfo{year}{2003}).
\newblock


\bibitem[Gonz{\'a}lez et~al\mbox{.}(2022)]%
        {gonzalez2022atmospheric}
\bibfield{author}{\bibinfo{person}{Jorge~L{\'o}pez Gonz{\'a}lez}, \bibinfo{person}{Theodore Chapman}, \bibinfo{person}{Kathryn Chen}, \bibinfo{person}{Hannah Nguyen}, \bibinfo{person}{Logan Chambers}, \bibinfo{person}{Seraj~AM Mostafa}, \bibinfo{person}{Jianwu Wang}, \bibinfo{person}{Sanjay Purushotham}, \bibinfo{person}{Chenxi Wang}, {and} \bibinfo{person}{Jia Yue}.} \bibinfo{year}{2022}\natexlab{}.
\newblock \showarticletitle{Atmospheric Gravity Wave Detection Using Transfer Learning Techniques}. In \bibinfo{booktitle}{\emph{2022 IEEE/ACM International Conference on Big Data Computing, Applications and Technologies (BDCAT)}}. IEEE, \bibinfo{pages}{128--137}.
\newblock


\bibitem[Guo et~al\mbox{.}(2007)]%
        {guo2007semantic}
\bibfield{author}{\bibinfo{person}{Dihua Guo}, \bibinfo{person}{Hui Xiong}, \bibinfo{person}{Vijay Atluri}, {and} \bibinfo{person}{Nabil Adam}.} \bibinfo{year}{2007}\natexlab{}.
\newblock \showarticletitle{Semantic feature selection for object discovery in high-resolution remote sensing imagery}. In \bibinfo{booktitle}{\emph{Advances in Knowledge Discovery and Data Mining: 11th Pacific-Asia Conference, PAKDD 2007, Nanjing, China, May 22-25, 2007. Proceedings 11}}. Springer, \bibinfo{pages}{71--83}.
\newblock


\bibitem[Hasan et~al\mbox{.}(2025)]%
        {hasan2025comparison}
\bibfield{author}{\bibinfo{person}{MD~Badrul Hasan}, \bibinfo{person}{Meilin Yu}, {and} \bibinfo{person}{Tim Oates}.} \bibinfo{year}{2025}\natexlab{}.
\newblock \showarticletitle{Comparison of Several Neural Network-Enhanced Sub-Grid Scale Stress Models for Meso-Scale Hurricane Boundary Layer Flow Simulation}. In \bibinfo{booktitle}{\emph{AIAA SCITECH 2025 Forum}}. \bibinfo{pages}{2212}.
\newblock


\bibitem[He et~al\mbox{.}(2016)]%
        {he2016deep}
\bibfield{author}{\bibinfo{person}{Kaiming He}, \bibinfo{person}{Xiangyu Zhang}, \bibinfo{person}{Shaoqing Ren}, {and} \bibinfo{person}{Jian Sun}.} \bibinfo{year}{2016}\natexlab{}.
\newblock \showarticletitle{Deep residual learning for image recognition}. In \bibinfo{booktitle}{\emph{Proceedings of the IEEE conference on computer vision and pattern recognition}}. \bibinfo{pages}{770--778}.
\newblock


\bibitem[Hozumi et~al\mbox{.}(2019)]%
        {hozumi2019geographical}
\bibfield{author}{\bibinfo{person}{Yuta Hozumi}, \bibinfo{person}{Akinori Saito}, \bibinfo{person}{Takeshi Sakanoi}, \bibinfo{person}{Atsushi Yamazaki}, \bibinfo{person}{Keisuke Hosokawa}, {and} \bibinfo{person}{Takuji Nakamura}.} \bibinfo{year}{2019}\natexlab{}.
\newblock \showarticletitle{Geographical and seasonal variability of mesospheric bores observed from the International Space Station}.
\newblock \bibinfo{journal}{\emph{Journal of Geophysical Research: Space Physics}} \bibinfo{volume}{124}, \bibinfo{number}{5} (\bibinfo{year}{2019}), \bibinfo{pages}{3775--3785}.
\newblock


\bibitem[Hu et~al\mbox{.}(2019)]%
        {hu2019measuring}
\bibfield{author}{\bibinfo{person}{Shensen Hu}, \bibinfo{person}{Shuo Ma}, \bibinfo{person}{Wei Yan}, \bibinfo{person}{Neil~P Hindley}, {and} \bibinfo{person}{Xianbin Zhao}.} \bibinfo{year}{2019}\natexlab{}.
\newblock \showarticletitle{Measuring internal solitary wave parameters based on VIIRS/DNB data}.
\newblock \bibinfo{journal}{\emph{International Journal of Remote Sensing}} \bibinfo{volume}{40}, \bibinfo{number}{20} (\bibinfo{year}{2019}), \bibinfo{pages}{7805--7816}.
\newblock


\bibitem[Jocher et~al\mbox{.}(2020)]%
        {glenn_jocher_2020_4154370}
\bibfield{author}{\bibinfo{person}{Glenn Jocher}, \bibinfo{person}{Alex Stoken}, \bibinfo{person}{Jirka Borovec}, \bibinfo{person}{NanoCode012}, \bibinfo{person}{ChristopherSTAN}, \bibinfo{person}{Liu Changyu}, \bibinfo{person}{Laughing}, \bibinfo{person}{tkianai}, \bibinfo{person}{Adam Hogan}, \bibinfo{person}{lorenzomammana}, \bibinfo{person}{yxNONG}, \bibinfo{person}{AlexWang1900}, \bibinfo{person}{Laurentiu Diaconu}, \bibinfo{person}{Marc}, \bibinfo{person}{wanghaoyang0106}, \bibinfo{person}{ml5ah}, \bibinfo{person}{Doug}, \bibinfo{person}{Francisco Ingham}, \bibinfo{person}{Frederik}, \bibinfo{person}{Guilhen}, \bibinfo{person}{Hatovix}, \bibinfo{person}{Jake Poznanski}, \bibinfo{person}{Jiacong Fang}, \bibinfo{person}{Lijun Yu}, \bibinfo{person}{changyu98}, \bibinfo{person}{Mingyu Wang}, \bibinfo{person}{Naman Gupta}, \bibinfo{person}{Osama Akhtar}, \bibinfo{person}{PetrDvoracek}, {and} \bibinfo{person}{Prashant Rai}.} \bibinfo{year}{2020}\natexlab{}.
\newblock \bibinfo{title}{{ultralytics/yolov5: v3.1 - Bug Fixes and Performance Improvements}}.
\newblock
\href{https://doi.org/10.5281/zenodo.4154370}{doi:\nolinkurl{10.5281/zenodo.4154370}}


\bibitem[Jovanovic(2018)]%
        {jovanovic2018nature}
\bibfield{author}{\bibinfo{person}{G Jovanovic}.} \bibinfo{year}{2018}\natexlab{}.
\newblock \showarticletitle{About the nature of gravitational and gravity waves}.
\newblock \bibinfo{journal}{\emph{Physics \& Astronomy International Journal}} \bibinfo{volume}{2}, \bibinfo{number}{2} (\bibinfo{year}{2018}), \bibinfo{pages}{75--77}.
\newblock


\bibitem[Li et~al\mbox{.}(2021)]%
        {li2021cascaded}
\bibfield{author}{\bibinfo{person}{Zhiqiang Li}, \bibinfo{person}{Xi Chen}, \bibinfo{person}{Jie Jiang}, \bibinfo{person}{Zhen Han}, \bibinfo{person}{Zhihong Li}, \bibinfo{person}{Tao Fang}, \bibinfo{person}{Hong Huo}, \bibinfo{person}{Qingli Li}, {and} \bibinfo{person}{Min Liu}.} \bibinfo{year}{2021}\natexlab{}.
\newblock \showarticletitle{Cascaded multiscale structure with self-smoothing atrous convolution for semantic segmentation}.
\newblock \bibinfo{journal}{\emph{IEEE Transactions on Geoscience and Remote Sensing}}  \bibinfo{volume}{60} (\bibinfo{year}{2021}), \bibinfo{pages}{1--13}.
\newblock


\bibitem[Liu et~al\mbox{.}(2019)]%
        {liu2019d}
\bibfield{author}{\bibinfo{person}{Zhiqun Liu}, \bibinfo{person}{Ruyi Feng}, \bibinfo{person}{Lizhe Wang}, \bibinfo{person}{Yanfei Zhong}, {and} \bibinfo{person}{Liqin Cao}.} \bibinfo{year}{2019}\natexlab{}.
\newblock \showarticletitle{D-Resunet: Resunet and dilated convolution for high resolution satellite imagery road extraction}. In \bibinfo{booktitle}{\emph{IGARSS 2019-2019 IEEE International Geoscience and Remote Sensing Symposium}}. IEEE, \bibinfo{pages}{3927--3930}.
\newblock


\bibitem[Miller et~al\mbox{.}(2013)]%
        {miller2013illuminating}
\bibfield{author}{\bibinfo{person}{Steven~D Miller}, \bibinfo{person}{William Straka~III}, \bibinfo{person}{Stephen~P Mills}, \bibinfo{person}{Christopher~D Elvidge}, \bibinfo{person}{Thomas~F Lee}, \bibinfo{person}{Jeremy Solbrig}, \bibinfo{person}{Andi Walther}, \bibinfo{person}{Andrew~K Heidinger}, {and} \bibinfo{person}{Stephanie~C Weiss}.} \bibinfo{year}{2013}\natexlab{}.
\newblock \showarticletitle{Illuminating the capabilities of the suomi national polar-orbiting partnership (NPP) visible infrared imaging radiometer suite (VIIRS) day/night band}.
\newblock \bibinfo{journal}{\emph{Remote Sensing}} \bibinfo{volume}{5}, \bibinfo{number}{12} (\bibinfo{year}{2013}), \bibinfo{pages}{6717--6766}.
\newblock


\bibitem[Miller et~al\mbox{.}(2015)]%
        {miller2015upper}
\bibfield{author}{\bibinfo{person}{Steven~D Miller}, \bibinfo{person}{William~C Straka~III}, \bibinfo{person}{Jia Yue}, \bibinfo{person}{Steven~M Smith}, \bibinfo{person}{M~Joan Alexander}, \bibinfo{person}{Lars Hoffmann}, \bibinfo{person}{Martin Setv{\'a}k}, {and} \bibinfo{person}{Philip~T Partain}.} \bibinfo{year}{2015}\natexlab{}.
\newblock \showarticletitle{Upper atmospheric gravity wave details revealed in nightglow satellite imagery}.
\newblock \bibinfo{journal}{\emph{Proceedings of the National Academy of Sciences}} \bibinfo{volume}{112}, \bibinfo{number}{49} (\bibinfo{year}{2015}), \bibinfo{pages}{E6728--E6735}.
\newblock


\bibitem[Mostafa et~al\mbox{.}(2025)]%
        {mostafa2025gwavenet}
\bibfield{author}{\bibinfo{person}{Seraj Al~Mahmud Mostafa}, \bibinfo{person}{Omar Faruque}, \bibinfo{person}{Chenxi Wang}, \bibinfo{person}{Jia Yue}, \bibinfo{person}{Sanjay Purushotham}, {and} \bibinfo{person}{Jianwu Wang}.} \bibinfo{year}{2025}\natexlab{}.
\newblock \showarticletitle{gWaveNet: Classification of Gravity Waves from Noisy Satellite Data Using Custom Kernel Integrated Deep Learning Method}. In \bibinfo{booktitle}{\emph{International Conference on Pattern Recognition}}. Springer, \bibinfo{pages}{164--180}.
\newblock


\bibitem[Mostafa et~al\mbox{.}(2023)]%
        {mostafa2023cnn}
\bibfield{author}{\bibinfo{person}{Seraj Al~Mahmud Mostafa}, \bibinfo{person}{Jinbo Wang}, \bibinfo{person}{Benjamin Holt}, \bibinfo{person}{Sanjay Purushotham}, {and} \bibinfo{person}{Jianwu Wang}.} \bibinfo{year}{2023}\natexlab{}.
\newblock \showarticletitle{CNN based Ocean Eddy Detection Using Cloud Services}. In \bibinfo{booktitle}{\emph{IGARSS 2023-2023 IEEE International Geoscience and Remote Sensing Symposium}}. IEEE, \bibinfo{pages}{4052--4055}.
\newblock


\bibitem[Murphy et~al\mbox{.}(2006)]%
        {murphy2006visible}
\bibfield{author}{\bibinfo{person}{RE Murphy}, \bibinfo{person}{Phillip Ardanuy}, \bibinfo{person}{Frank~J Deluccia}, \bibinfo{person}{JE Clement}, {and} \bibinfo{person}{Carl~F Schueler}.} \bibinfo{year}{2006}\natexlab{}.
\newblock \showarticletitle{The {Visible Infrared Imaging Radiometer Suite}}.
\newblock In \bibinfo{booktitle}{\emph{Earth Science Satellite Remote Sensing: Vol. 1: Science and Instruments}}. \bibinfo{publisher}{Springer}, \bibinfo{pages}{199--223}.
\newblock


\bibitem[{NOAA Gravity Wave Data}({[n.\,d.]})]%
        {gravitywavedata}
\bibfield{author}{\bibinfo{person}{{NOAA Gravity Wave Data}}.} \bibinfo{year}{[n.\,d.]}\natexlab{}.
\newblock \bibinfo{title}{Gravity Wave VIIRS data source}.
\newblock
\urldef\tempurl%
\url{https://noaa-nesdis-n20-pds.s3.amazonaws.com/index.html#reprocessed/}
\showURL{%
\tempurl}


\bibitem[Park and Paik(2022)]%
        {park2022pyramid}
\bibfield{author}{\bibinfo{person}{Hyeokjin Park} {and} \bibinfo{person}{Joonki Paik}.} \bibinfo{year}{2022}\natexlab{}.
\newblock \showarticletitle{Pyramid attention upsampling module for object detection}.
\newblock \bibinfo{journal}{\emph{IEEE Access}}  \bibinfo{volume}{10} (\bibinfo{year}{2022}), \bibinfo{pages}{38742--38749}.
\newblock


\bibitem[Qiu et~al\mbox{.}(2022)]%
        {qiu2022a2sppnet}
\bibfield{author}{\bibinfo{person}{Yu Qiu}, \bibinfo{person}{Yun Liu}, \bibinfo{person}{Yanan Chen}, \bibinfo{person}{Jianwen Zhang}, \bibinfo{person}{Jinchao Zhu}, {and} \bibinfo{person}{Jing Xu}.} \bibinfo{year}{2022}\natexlab{}.
\newblock \showarticletitle{A2SPPNet: Attentive atrous spatial pyramid pooling network for salient object detection}.
\newblock \bibinfo{journal}{\emph{IEEE Transactions on Multimedia}}  \bibinfo{volume}{25} (\bibinfo{year}{2022}), \bibinfo{pages}{1991--2006}.
\newblock


\bibitem[Tian et~al\mbox{.}(2025)]%
        {tian2025yolov12}
\bibfield{author}{\bibinfo{person}{Yunjie Tian}, \bibinfo{person}{Qixiang Ye}, {and} \bibinfo{person}{David Doermann}.} \bibinfo{year}{2025}\natexlab{}.
\newblock \showarticletitle{YOLOv12: Attention-Centric Real-Time Object Detectors}.
\newblock \bibinfo{journal}{\emph{arXiv preprint arXiv:2502.12524}} (\bibinfo{year}{2025}).
\newblock


\bibitem[Tushar et~al\mbox{.}(2024)]%
        {tushar2024cloudunet}
\bibfield{author}{\bibinfo{person}{Zahid~Hassan Tushar}, \bibinfo{person}{Adeleke Ademakinwa}, \bibinfo{person}{Jianwu Wang}, \bibinfo{person}{Zhibo Zhang}, {and} \bibinfo{person}{Sanjay Purushotham}.} \bibinfo{year}{2024}\natexlab{}.
\newblock \showarticletitle{Cloudunet: Adapting unet for retrieving cloud properties}. In \bibinfo{booktitle}{\emph{IGARSS 2024-2024 IEEE International Geoscience and Remote Sensing Symposium}}. IEEE, \bibinfo{pages}{7163--7167}.
\newblock


\bibitem[Vaswani(2017)]%
        {vaswani2017attention}
\bibfield{author}{\bibinfo{person}{Ashish Vaswani}.} \bibinfo{year}{2017}\natexlab{}.
\newblock \showarticletitle{Attention is all you need}.
\newblock \bibinfo{journal}{\emph{arXiv preprint arXiv:1706.03762}} (\bibinfo{year}{2017}).
\newblock


\bibitem[Wang et~al\mbox{.}(2019)]%
        {wang2019salient}
\bibfield{author}{\bibinfo{person}{Wenguan Wang}, \bibinfo{person}{Shuyang Zhao}, \bibinfo{person}{Jianbing Shen}, \bibinfo{person}{Steven~CH Hoi}, {and} \bibinfo{person}{Ali Borji}.} \bibinfo{year}{2019}\natexlab{}.
\newblock \showarticletitle{Salient object detection with pyramid attention and salient edges}. In \bibinfo{booktitle}{\emph{Proceedings of the IEEE/CVF conference on computer vision and pattern recognition}}. \bibinfo{pages}{1448--1457}.
\newblock


\bibitem[Wang and Ji(2018)]%
        {wang2018smoothed}
\bibfield{author}{\bibinfo{person}{Zhengyang Wang} {and} \bibinfo{person}{Shuiwang Ji}.} \bibinfo{year}{2018}\natexlab{}.
\newblock \showarticletitle{Smoothed dilated convolutions for improved dense prediction}. In \bibinfo{booktitle}{\emph{Proceedings of the 24th ACM SIGKDD International Conference on Knowledge Discovery \& Data Mining}}. \bibinfo{pages}{2486--2495}.
\newblock


\bibitem[Woo et~al\mbox{.}(2018)]%
        {woo2018cbam}
\bibfield{author}{\bibinfo{person}{Sanghyun Woo}, \bibinfo{person}{Jongchan Park}, \bibinfo{person}{Joon-Young Lee}, {and} \bibinfo{person}{In~So Kweon}.} \bibinfo{year}{2018}\natexlab{}.
\newblock \showarticletitle{Cbam: Convolutional block attention module}. In \bibinfo{booktitle}{\emph{Proceedings of the European conference on computer vision (ECCV)}}. \bibinfo{pages}{3--19}.
\newblock


\bibitem[Zhang et~al\mbox{.}(2020)]%
        {zhang2020multiple}
\bibfield{author}{\bibinfo{person}{Lun Zhang}, \bibinfo{person}{Junhua Zhang}, \bibinfo{person}{Zonggui Li}, {and} \bibinfo{person}{Yingchao Song}.} \bibinfo{year}{2020}\natexlab{}.
\newblock \showarticletitle{A multiple-channel and atrous convolution network for ultrasound image segmentation}.
\newblock \bibinfo{journal}{\emph{Medical Physics}} \bibinfo{volume}{47}, \bibinfo{number}{12} (\bibinfo{year}{2020}), \bibinfo{pages}{6270--6285}.
\newblock


\bibitem[Zhou et~al\mbox{.}(2020)]%
        {zhou2020scale}
\bibfield{author}{\bibinfo{person}{Feng Zhou}, \bibinfo{person}{Yong Hu}, {and} \bibinfo{person}{Xukun Shen}.} \bibinfo{year}{2020}\natexlab{}.
\newblock \showarticletitle{Scale-aware spatial pyramid pooling with both encoder-mask and scale-attention for semantic segmentation}.
\newblock \bibinfo{journal}{\emph{Neurocomputing}}  \bibinfo{volume}{383} (\bibinfo{year}{2020}), \bibinfo{pages}{174--182}.
\newblock


\bibitem[Zhou et~al\mbox{.}(2018)]%
        {zhou2018d}
\bibfield{author}{\bibinfo{person}{Lichen Zhou}, \bibinfo{person}{Chuang Zhang}, {and} \bibinfo{person}{Ming Wu}.} \bibinfo{year}{2018}\natexlab{}.
\newblock \showarticletitle{D-LinkNet: LinkNet with pretrained encoder and dilated convolution for high resolution satellite imagery road extraction}. In \bibinfo{booktitle}{\emph{Proceedings of the IEEE conference on computer vision and pattern recognition workshops}}. \bibinfo{pages}{182--186}.
\newblock


\end{thebibliography}
\end{document}